\documentclass[prb,twocolumn,showpacs,groupedaddress]{revtex4}

\usepackage{epsfig,color}
\usepackage{dcolumn}% Align table columns on decimal point

\begin{document}

\title{Theory of magnetic excitations in iron-based layered superconductors}

\author {M.M.~Korshunov$^{1,2}$}
 \email{maxim@mpipks-dresden.mpg.de}
\author {I.~Eremin$^{1,3}$}
\affiliation {$^1$Max-Planck-Institut f\"{u}r Physik komplexer
Systeme, D-01187 Dresden, Germany}

\affiliation {$^2$L.V. Kirensky Institute of Physics, Siberian
Branch of Russian Academy of Sciences, 660036 Krasnoyarsk, Russia}

\affiliation {$^3$Institute f\"{u}r Mathematische und Theoretische
Physik, TU Braunschweig, D-38106 Braunschweig, Germany}

\date{April 10, 2008}

\begin{abstract}
Based on the effective four-band model we analyze the spin response in the
normal and superconducting states of the Fe-pnictide superconductors. While the
normal state spin excitations are dominated by the continuum of the
interorbital antiferromagnetic fluctuations and the intraband spin density wave
fluctuations, the unconventional superconductivity yields different feedback.
The resonance peak in form of the well-defined spin exciton occurs {\it only}
for the interband scattering at the antiferromagnetic momentum ${\bf Q}_{AFM}$
for the $s_\pm$ (extended $s$-wave) superconducting order parameter and it
disappears rapidly for ${\bf q} < {\bf Q}_{AFM}$. The resonance feature is
extremely weak for the $d_{x^2 -y^2}$-wave order parameter due to specific
Fermi surface topology of these compounds. The essential difference between
$s_\pm$-wave and $d_{x^2 -y^2}$-wave symmetries for the magnetic excitations
can be used for experimental determination of the superconducting wave function
symmetry.
\end{abstract}


\maketitle

The relation between unconventional superconductivity and magnetism is one of
the most interesting topics in the condensed matter physics. In contrast to the
usual electron-phonon mediated superconductors where the paramagnetic spin
excitations are suppressed below superconducting transition temperature due to
the formation of the Cooper pairs with total spin $S=0$, in unconventional
superconductors, such as layered cuprates or heavy fermion superconductors, a
bound state (spin resonance) with a high intensity forms below T$_c$
\cite{rossat,sato,broholm}. The fact that the superconducting gap is changing
sign at a different parts of the Fermi surface together with a presence of the
strong electronic correlations yields such an enhancement of the spin response
\cite{eschrig}. Most interestingly, an observation of the resonance peak
indicates not only that Cooper-pairing is unconventional but also that the
magnetic fluctuations are most relevant for superconductivity \cite{chubukov}.

Since the discovery of superconductivity in the quaternary oxypnictides LaFePO
\cite{Kamihara} and LaNiPO \cite{Watanabe}, a new class of high-T$_c$ materials
with Fe-based layered structure is emerging
\cite{Kamihara2,Chen,Yang,Wen,Chen2,Ren,Chen3}. Although the microscopic nature
of superconductivity in these compounds remains unclear at present, certain
aspects have been already discussed
\cite{Lebegue,Singh,Boeri,Mazin,Ma,Kuroki,Haule,Xu,Dong,Dai,Han,Li,Raghu}. In
particular, \textit{ab initio} band structure calculations
\cite{Lebegue,Singh,Boeri,Mazin,Ma,Kuroki} have shown that the conductivity and
superconductivity in these systems are associated with the Fe-pnictide layer,
and the electronic density of states (DOS) near the Fermi level shows maximum
contribution from the Fe-$3d$ orbitals. The resulting Fermi surface consists of
two hole (\textit{h}) and two electron (\textit{e}) pockets. The normal state
magnetic spin susceptibility determined from these bands \cite{Xu} exhibits
both small ${\bf q} \sim 0$ fluctuations and antiferromagnetic commensurate
spin density wave (SDW) peaks.

In this Rapid Communication, using the four-band tight-binding model we study
theoretically the spin response in the normal and superconducting states of
Fe-based pnictide superconductors. We show that the resulting magnetic
fluctuation spectrum calculated within random-phase approximation (RPA)
consists of two contributions. The first one is from the antiferromagnetic
(AFM) spin fluctuations peaked at ${\bf Q}_{AFM}=(\pi,\pi)$ arising due to the
interband scattering. The second contribution comes from the intraband
scattering and results in a broad continuum of the SDW fluctuations with a
small momenta. We show that the unconventional superconductivity yields
different feedback on the magnetic excitation spectrum. The resonance peak in
form of the spin exciton occurs only for the interband scattering at the AFM
momentum for the $s_\pm$-wave superconducting order parameter. This peak is
confined to the AFM wave vector and disappears rapidly away from it. We suggest
that the superconductivity is most likely $s_\pm$-wave and is driven by the
repulsive interaction.

The Fe ions form a square lattice in the FeAs layer of LaFeAsO system, which is
interlaced with the second square lattice of As ions. Due to the fact that As
ions sit in the center of each square plaquette of the Fe lattice and are
displaced above and below the Fe plane, the crystallographic unit cell contains
two Fe and two As ions. The band structure calculations
\cite{Lebegue,Singh,Boeri,Mazin,Ma,Kuroki} show that three Fe-$3d$ states
($d_{xz}$, $d_{yz}$, and $d_{xy}$) give the main contribution to the density of
states close to the Fermi level and that these states disperse weakly in the
$z$-direction. The resulting Fermi surface consists of two hole (h) pockets
centered around the $\Gamma=(0,0)$ point and two electron (e) pockets centered
around the $M=(\pi,\pi)$ point of the {\it folded} Brillouin zone (BZ)
\cite{Mazin}. Note, the {\it folded} BZ corresponds to the case of two Fe atoms
per unit cell, and the wave vector $(\pi,\pi)$ in the {\it folded} BZ
corresponds to the $(\pi,0)$ wave vector in the {\it unfolded} BZ (related to
the case of one Fe per unit cell). To model the resulting band structure we
assume the following single-electron Hamiltonian
\begin{eqnarray}
H_0 = - \sum\limits_{{\bf k},\alpha ,\sigma } {{\epsilon^i} n_{{\bf
k} i \sigma } } - \sum\limits_{{\bf k}, i, \sigma}  t_{{\bf k}}^{i}
d_{{\bf k} i \sigma }^\dag d_{{\bf k} i \sigma}, \label{eq:H0}
\end{eqnarray}
where $i=\alpha_1,\alpha_2,\beta_1, \beta_2$ refer to the band
indices, $\epsilon^i$ are the on-site single-electron energies,
$t_{{\bf k}}^{\alpha_{1},\alpha_{2}} = t^{\alpha_{1},\alpha_{2}}_1
\left(\cos k_x+\cos k_y \right)+ t^{\alpha_{1},\alpha_{2}} _2 \cos
k_x \cos k_y$ is the electronic dispersion that yields hole pockets
centered around the $\Gamma$ point, and $t_{{\bf
k}}^{\beta_{1},\beta_{2}} = t^{\beta_{1},\beta_{2}}_1 \left(\cos
k_x+\cos k_y \right)+ t^{\beta_{1},\beta_{2}}_2 \cos \frac {k_x}{2}
\cos \frac{k_y}{2}$ is the dispersion that results in the electron
pockets around the $M$ point. Using the abbreviation $(\epsilon^i,
t_1^i, t_2^i)$ we choose the parameters $(-0.60,0.30,0.24)$ and
$(-0.40,0.20,0.24)$ for the $\alpha_1$ and $\alpha_2$ bands,
respectively, and $(1.70,1.14,0.74)$  and $(1.70,1.14,-0.64)$  for
the $\beta_1$ and $\beta_2$ bands, correspondingly (all values are
in eV).

\begin{figure}
\includegraphics[angle=0,width=1.0\linewidth]{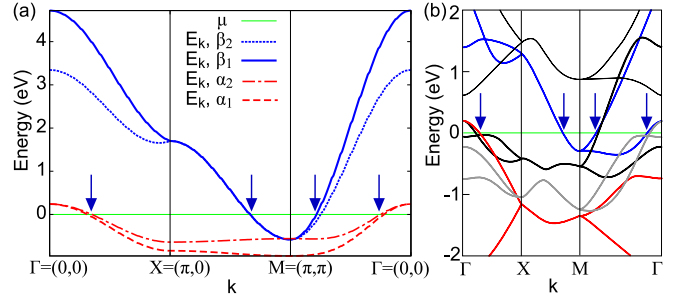}
\caption{(Color online) Calculated two-dimensional effective band
structure along the main symmetry directions of the {\it folded} BZ
(a) for the LaFeAsO system, and the reproduction (b) of the
corresponding LDA band structure
\cite{Lebegue,Singh,Boeri,Mazin,Kuroki}. The arrows in (a) and (b)
indicate the points where bands cross the Fermi level.
Note, the shade (color) of the curves in (b) is used just as a guide to the
eye and does not reflect the actual hybridization of the bands.}
\label{fig1}
\end{figure}

\begin{figure}
\includegraphics[angle=0,width=1.0\linewidth]{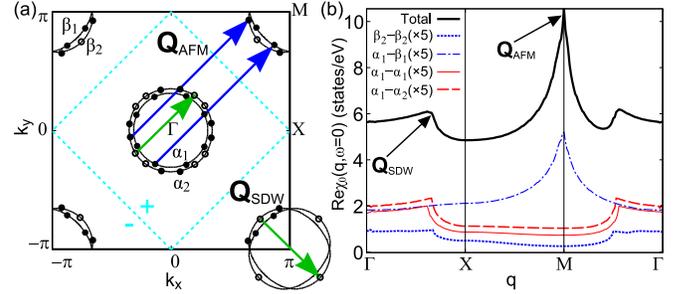}
\caption{(Color online) (a) Calculated Fermi surface topology for
the LaFeAsO system. The arrows indicate the main scattering wave
vectors. The filled dots refer to the states connected by the
interband scattering at the AFM wave vector ${\bf Q}_{AFM}$,
while the open dots denote the interband and intraband
scattering at the incommensurate wave vector ${\bf Q}_{SDW}$. The
dashed (cyan) lines and the $+$,$-$ signs depict the position of the
nodes and the corresponding phase of the $s_\pm$
superconducting order parameter. (b) Calculated real
part of the one-loop spin susceptibility along the main symmetry
directions of the first {\it folded} BZ. The thick solid (black) curve
refer to the total susceptibility while the other (red and blue) curves
refer to the partial contributions which are multiplied by a factor of 5
for the sake of the presentation.
The main scattering wave vectors shown in (a) are also indicated in (b).}
\label{fig2}
\end{figure}

In Figs.~\ref{fig1}(a) and \ref{fig2}(a) we show the resulting band structure
and the corresponding Fermi surface for the undoped case, $x=0$. The Fermi
surface consists of the two hole ($\alpha_1$ and $\alpha_2$) and two electron
($\beta_1$ and $\beta_2$) pockets. The $\beta$ bands show much broader
bandwidth and are degenerate along $X-M$ direction which is a consequence of
the hybridization of the underlaying $d_{xz}$ and $d_{yz}$ orbitals within the
{\it folded} BZ. The $\alpha$ bands centered around the $\Gamma$ point are
narrower which also results in the significant contribution to the density of
states. The chosen band structure reproduces correctly the local-density
approximation (LDA) Fermi surface topology and the corresponding values of the
Fermi velocities. In particular, we have selected the on-site energies and the
hopping matrix elements assuming the compensated metal at zero doping and the
filling factor $n=4$ (we further assume that there exists another band below
the Fermi level which is fully occupied and not considered here). Additionally,
we take into account the details of the electronic dispersions of the bands
which form the corresponding hole and electron Fermi surface pockets. The
comparison between our effective model and the \textit{ab initio} density
functional calculations \cite{Lebegue,Singh,Boeri,Mazin,Kuroki} can be seen
from Fig.~\ref{fig1}(b) where we display the electronic dispersion from
Ref.~\onlinecite{Lebegue}. Note, the hole Fermi surfaces shifted by $(\pi,\pi)$
is fully nested with that of the electron pockets which is also in full
agreement with LDA results. Here, the position of the chemical potential $\mu$
has been deduced from the equation $n=4+x$.
We note that although the Fermi surface obtained previously in the
effective two-band model \cite{Raghu} reproduces correctly the one
obtained within LDA calculations, the actual evolution of the
dispersion deviates significantly.

Next we consider the one-loop contribution to the spin susceptibility that
includes the intraband and the interband contributions:
\begin{eqnarray}
\chi_0^{ij}({\bf q},{\rm i} \omega_m)&=&- \frac{T}{2N}
\sum_{{\bf k}, \omega_n} {\rm Tr}
\left[ G^i({\bf k + q}, {\rm i} \omega_n + {\rm i} \omega_m) G^j({\bf k} , {\rm i} \omega_n) \right. \nonumber \\
&+& \left. F^i({\bf k + q}, {\rm i} \omega_n + {\rm i} \omega_m) F^j({\bf k} , {\rm i} \omega_n)\right]
\label{eq:bare_chi}
\end{eqnarray}
where $i$, $j$ again refer to the different band indices. $G^i$ and $F^i$ are
the normal and anomalous (superconducting) Green functions, respectively.

In Fig.~\ref{fig2} we present the results for the real part of the total
(physical) spin susceptibility $\chi_0 ({\bf q},{\rm i}
\omega_m)=\sum_{i,j}\chi^{i,j}_0({\bf q},{\rm i}\omega_m)$, as well as the
partial contributions. The total susceptibility is dominated by the
scattering at the AFM wave vector ${\bf Q}_{AFM}$ which is originated due to
the interband ($\alpha \to \beta$) scattering. It is interesting to note that
the intraband and interband scattering within $\alpha$ and $\beta$ bands are
very similar and are responsible for the broad hump around the ${\bf Q}_{SDW}$
wave vector.

In the following we shall discuss the possible influence of the
superconductivity driven by the short-range magnetic or charge fluctuations on
the magnetic susceptibility. It has been already argued that most likely the
superconductivity in these family of compounds is of unconventional origin and
is driven either by the interband AFM fluctuations or by the intraband SDW
fluctuations. However, one has to stress that even if the Cooper-pairing is
driven by the interband fluctuations it still refers to the two fermionic
states on the very same $\alpha$ or $\beta$ bands. The standard Cooper-pairing
for the two fermions from the different bands will be suppressed, since there
are no states with ${\bf k}$ and ${\bf -k}$ that can be connected at the
different Fermi surfaces by the AFM momentum, as could easily be seen in
Fig.~\ref{fig2}(a). Therefore, we expect that inter-orbital AFM fluctuations
will drive superconductivity in the $\alpha$ and $\beta$ bands. The latter
should also result in the very same value of the superconducting gap in both
bands. The repulsive nature of the interaction would then require
\cite{mazinyakov} the superconducting gap that satisfies $\Delta^i_{\bf k} =
-\Delta^j_{{\bf k+Q}_{AFM}}$. Thus, we consider the magnetic susceptibility in
the superconducting state assuming $d_{x^2 - y^2}$-wave [$\Delta_{\bf k} =
\frac{\Delta_0}{2}\left( \cos k_x - \cos k_y \right)$] and $s_\pm$-wave
[$\Delta_{\bf k} = \frac{\Delta_0}{2}\left( \cos k_x + \cos k_y \right)$]
symmetries of the order parameter which both satisfy the condition given above.

For the four-band model considered here the effective interaction will consist
of the on-site Hubbard intraband repulsion $U$ and the Hund's coupling $J$.
There is also an interband Hubbard repulsion $U'$, which however does not
contribute to the RPA susceptibility. Within RPA the spin response has a matrix
form:
\begin{eqnarray}
\hat{\chi}_{RPA}({\bf q},{\rm i}\omega_m)=\left[\mathbf{I}-{\bf
\Gamma} \hat{\chi}_0({\bf q},{\rm i}\omega_m)\right]^{-1}
\hat{\chi}_0({\bf q},{\rm i}\omega_m) \label{eq:chi_RPA}
\end{eqnarray}
where ${\bf I}$ is a unit matrix and $\hat{\chi}_0({\bf q},{\rm i}\omega_m)$ is
$4 \times 4$ matrix formed by the interband and intraband bare susceptibilities
determined by Eq.~(\ref{eq:bare_chi}). The vertex is given by
\begin{eqnarray}
{\bf \Gamma} = \left[\begin{array}{cccc} U & J/2 & J/2 & J/2
\\ J/2 & U & J/2 & J/2 \\ J/2 & J/2 & U & J/2 \\ J/2 & J/2 & J/2 & U
\end{array}\right],
\label{eq_coupling}
\end{eqnarray}
and we assume here $J=0.2 U$ and $U \sim t^{\beta_1}_1$. Note that the value of
$U$ was chosen in order to stay in the paramagnetic phase. We have to note that
our interaction parameters are carrying the band indices. Therefore, we neglect
the possible orbital correlations. Whether this may play an important role
needs to be addressed carefully and is beyond the scope of the present study.
Though the current experimental and theoretical belief is such that the orbital
physics is not involved in the physics of ferropnictides due to strong
hybridization of all $d$-orbitals.

\begin{figure}
\includegraphics[angle=0,width=1.0\linewidth]{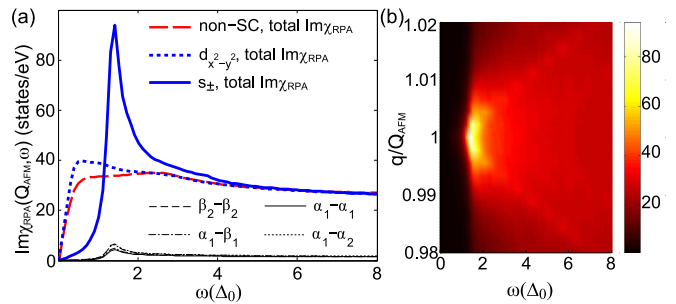}
\caption{(Color online) (a) Calculated imaginary part of the RPA
spin susceptibility at the AFM wave vector ${\bf Q}_{AFM}$ as a
function of frequency in the normal and superconducting states. The
thick dashed (red), dotted (blue), and solid (blue) curves
correspond to the total RPA susceptibility. The thin (black) curves
refer to the partial RPA contributions for the interband and
intraband transitions in the $s_\pm$ superconducting state. (b)
Calculated imaginary part of the total RPA spin susceptibility in
the $s_\pm$ state as a function of frequency and momentum along (1,1) direction.
For the numerical purposes we set the damping constant $\delta^+ = 0.8$ meV.}
\label{fig3}
\end{figure}

In Fig.~\ref{fig3}(a) we show the results for the total RPA susceptibility,
$\chi_{RPA} ({\bf q},{\rm i} \omega_m)=\sum_{i,j}\chi^{i,j}_{RPA}({\bf q},{\rm
i}\omega_m)$, as a function of frequency at the AFM momentum ${\bf Q}_{AFM}$.
One finds that in the normal state the spin response does not show a
well-defined peak but rather a broad continuum of the spin fluctuations. The
origin for this is that the RPA enhancement of the AFM spin fluctuations is
determined by the $\det| \mathbf{I}-{\bf \Gamma} \hat{\chi}({\bf q},{\rm
i}\omega_m)|$. One has to remember that the intraband on-site Coulomb repulsion
$U$ will strengthen the corresponding intraband fluctuations and the Hund's
exchange will only increase directly the instability towards interorbital AFM
fluctuations. Given the fact that each of the bare susceptibilities slightly
differ from band to band as shown in Fig.~\ref{fig2}(b), the RPA does not yield
a well-defined pole. Thus one obtains simply a continuum of the fluctuations.
The situation changes in the superconducting state. The quasiparticles at the
Fermi surface of the $\alpha$ and $\beta$ bands connected by the AFM wave
vector possess the condition $\Delta_{\bf k} = - \Delta_{{\bf k+Q}_{AFM}}$ for
the $s_\pm$ order parameter. The imaginary part of the interband magnetic
susceptibility is zero for small frequencies due to the opening of the gap, and
then it experiences a discontinuous jump at $\Omega_c = \min \left(
|\Delta_{\bf k}| + |\Delta_{{ \bf k+Q}_{AFM}}| \right)$. Correspondingly, the
real part of the interband ($\alpha \to \beta$) susceptibility will show the
logarithmic singularity. This fulfils the resonance condition for the interband
susceptibility: $1-(J/2){\rm Re} \chi^{\alpha \beta}_0 ({\bf
Q}_{AFM},\omega_{res}) = 0$ and ${\rm Im}\chi^{\alpha \beta}_0 ({\bf
Q}_{AFM},\omega_{res}) = 0$. Moreover, the intraband bare susceptibilities are
small at this wave vector due to the direct gap, {\it i.e.} no states at the
Fermi level can be connected by the ${\bf Q}_{AFM}$ for the intraband
transitions. Therefore, a single resonant pole will occur for all components of
the RPA spin susceptibility at $\omega_{res} \leq \Omega_c$ and the spin
excition will form. This is evidently seen from Fig.~\ref{fig3}(a). Due to the
single pole in the denominator all components of the RPA susceptibilities
behave very similarly and the total susceptibility shows a well-defined
resonance peak.

In the case of $d_{x^2-y^2}$ superconducting gap the situation is more
complicated. As clearly seen from Fig.~\ref{fig2}(a), the AFM wave vector
connects states rather close to the node of the $d_{x^2-y^2}$ superconducting
order parameter and the overall gap in ${\rm Im}\chi^{\alpha \beta}_{0}$
determined by $\Omega_c$ is small. At the same time even for this symmetry the
resonance condition can be fulfilled due to the fact that $\Delta_{\bf k} = -
\Delta_{{\bf k+Q}_{AFM}}$. However, because of the smallness of $\Omega_c \ll
\Delta_0$ the total RPA susceptibility shows a moderate enhancement with
respect to the normal state value, as seen in Fig.~\ref{fig3}(a). Therefore,
the resonance peak is pronounced only for the $s_\pm$ order parameter. Such a
distinct behavior for the two various order parameters can be clearly resolved
by the inelastic neutron scattering experiments and therefore can be a direct
tool to clarify the symmetry of the superconducting order parameter in these
systems. Like for $d_{x^2-y^2}$ case, we have also found that there is no spin
resonance for $d_{xy}$- and $d_{x^2-y^2}+{\rm i}d_{xy}$-wave symmetries (due to
their similarity we do not present these results).

Finally we address the evolution of the resonance peak away from the AFM wave
vector. In Fig.~\ref{fig3}(b) we show the total RPA susceptibility as a
function of the momentum and frequency. Note that the $s_\pm$ superconducting
gap changes only slightly at the $\alpha$ and $\beta$ Fermi surfaces and can be
considered nearly as a constant. Therefore, one always finds $\Delta_{\bf k} =
- \Delta_{{\bf k+q}_n}$ as long as the wave vector ${\bf q}_n < {\bf Q}_{AFM}$
connects the states at the Fermi surface of one of the $\alpha$ and one of the
$\beta$ bands. However, as it is also clearly seen from Fig.~\ref{fig2}(b) the
nesting condition is very sensitive to the variation of ${\bf q}_n$ away from
${\bf Q}_{AFM}$. Therefore, already at ${\bf q}_n \approx 0.995 {\bf Q}_{AFM}$
the ${\rm Re} \chi^{\alpha \beta}_0 ({\bf q}_n,\omega_{res})$ is much smaller
than its value at ${\bf Q}_{AFM}$. As a result the resonance peak is confined
to the AFM momentum and does not disperse as it occurs for example in
high-T$_c$ cuprates.

In conclusion, we have analyzed the behavior of the magnetic spin
susceptibility in Fe-pnictide superconductors. We show that the magnetic
fluctuation spectrum calculated within RPA consists of (i) the continuum of the
AFM spin fluctuations peaked at ${\bf Q}_{AFM} = (\pi,\pi)$ that arise due to
the interband scattering, and (ii) a low-${\bf q}$ fluctuations around the
${\bf Q}_{SDW}$ due to the intraband scattering. We show that the
unconventional superconductivity yields different feedback on the magnetic
excitation spectrum. The resonance peak in form of the spin exciton occurs only
for the interband scattering at the AFM momentum for the $s_\pm$
superconducting order parameter. We also find that the resonance peak is
confined to the AFM wave vector and disappears rapidly for ${\bf q} < {\bf
Q}_{AFM}$.

\textit{Note added.} After submission of our paper, we became aware of the
experimental observation of the predicted resonance \cite{Christianson} and of
the study by Meier and Scalapino \cite{meier} who reached some similar
conclusions as ours.

We would like to thank A. Donkov, D. Parker, and P. Thalmeier for useful
discussions. I.E. acknowledges support from Volkswagen Foundation.

\end{document}